\documentclass[10pt,conference]{IEEEtran}

\newtheorem{theorem}{Theorem}
\newtheorem{claim}{Claim}
\usepackage{amsmath,verbatim}
\usepackage{graphicx}
\usepackage{graphics}
\usepackage{epsfig}
\usepackage{tikz}

\DeclareMathSizes{10}{9}{6}{4}
\begin{document}

\title{Analysis of the Second Moment of the LT Decoder}

\author{\authorblockN{Ghid Maatouk and Amin Shokrollahi}
\authorblockA{EPFL\\
Lausanne, Switzerland\\
Email: \{ghid.maatouk,amin.shokrollahi\}@epfl.ch}
}

\maketitle

\begin{abstract}
We analyze the second moment of the ripple size during the LT decoding
process and prove that the standard deviation of the ripple size for
an LT-code with length $k$ is of the order of $\sqrt k.$ Together with
a result by Karp et. al stating that the expectation of the ripple
size is of the order of $k$ \cite{finite}, this gives bounds on the
error probability of the LT decoder. We also give an analytic
expression for the variance of the ripple size up to terms of constant
order, and refine the expression in \cite{finite} for the expectation
of the ripple size up to terms of the order of $1/k$, thus providing a
first step towards an analytic finite-length analysis of LT decoding.  
\end{abstract}

\section{Introduction}
We assume the reader is familiar with Fountain codes, LT-codes and
belief propagation (BP) decoding. For details, the reader is referred
to \cite{luby}, \cite{raptor}. 

We consider LT-codes with parameters $(k,\Omega(x))$, where $k$ is the
message length and $\Omega(x)=\sum \Omega_i x^i$ is the degree
distribution of the output symbols during encoding. 
An important set to consider is the set of output symbols of degree
$1$ (the \textit{ripple}). The size of the ripple varies during the
decoding process, as high-degree output symbols become of degree $1$
after the removal of their edges, and as ripple elements become
useless after the recovering of their unique neighbor. 

The decoding is in error if and only if the ripple becomes empty
before all the input symbols are recovered. A natural question is thus
whether we can track the size of the ripple, in the expectation,
during the decoding process. Karp et al. \cite{finite} proved that the
expected ripple size is linear in $k$ throughout most of the decoding
process. Their asymptotic analytic expressions for the expected ripple
size can be found in section \ref{prelim}. They also derive an
expression for the expected \textit{cloud} size throughout decoding,
where the cloud is defined at each decoding step as the set of output
symbols of degree strictly higher than $1$. 

In this paper, we extend their analysis in two ways. First, we
consider higher moments of the cloud and ripple size in order to upper
bound the error probability of the LT decoder. More specifically, we
use similar methods to derive an expression for the variance of the
ripple size and prove that it is also linear in $k$ throughout most of
the decoding process. We can then use this expression together with
the expression for the expectation to offer a guarantee for successful
decoding, as follows: if, for fixed LT-code parameters, $R(u)$ is the
expectation and $\sigma_R(u)$ is the standard deviation of the ripple
size when $u$ symbols are unrecovered, then if the function 
\begin{equation}\label{hc}
 h_c(u) = R(u) - c \cdot \sigma_R(u)
\end{equation}
for some parameter $c$ never takes negative values, we can upper
bound the error probability of the LT decoder by the probability that
the ripple size deviates from its mean by more than $c$ standard
deviations. 

Second, we take the first step towards an analytic finite-length
analysis of the LT decoder, by providing exact expressions for the
expectation (variance) of the ripple size up to $O(1/k)$ (constant)
terms. This is done by considering lower-order terms in the difference
equations, but also by getting tight bounds on the discrepancy
introduced by approximating difference equations by differential
equations. 

It is worthy to note that the expressions we deal with are valid for
``most of the decoding process,'' that is, the analysis breaks down
when the number of unrecovered symbols is no longer a constant
fraction of $k$. This is no issue, however, when one considers Raptor
codes, which need only a constant fraction of the input symbols to be
recovered by the LT decoder \cite{raptor}.

\section{Preliminaries - an expression for the expected ripple size}\label{prelim}
Let $u$ be the number of unrecovered (\textit{undecoded}) input
symbols at a given decoding step. Define the decoder to be in state
$(c,r,u)$ if the cloud size is $c$ and the ripple size is $r$ at this
decoding step. To each state $(c,r,u)$, we can associate the
probability $p_{c,r,u}$ of the decoder being in this state. Define the
\textit{state generating function} of the LT decoder when $u$ symbols
are undecoded as 
\begin{equation*}
P_u(x,y) = \sum_{c \geq 0, r \geq 1} p_{c,r,u} x^c y^{r-1}.
\end{equation*}
The following theorem by Karp et al. gives a recursion for the state
generating function of the LT decoder.\\ 

\begin{theorem}\cite{finite}
 Suppose that the original code has $k$ input symbols and that
 $n=k(1+\delta)$ output symbols have been collected for
 decoding. Further, denote by $\Omega_i,$ $i=2,\ldots,D,$ the
 probability that an output symbol is of degree $i,$ where $D$ is the
 maximum degree of an output symbol. Then we have for
 $u=k+1,k,\ldots,1$ 
\begin{equation}\label{recursion}
 \begin{split}
  P_{u-1}(x,y) &= \frac{1}{y}\bigg[P_u\left(x(1-p_u)+y p_u, \frac{1}{u}+y\left(1-\frac{1}{u}\right)\right)\\
&\qquad - P_u\left(x(1-p_u),\frac{1}{u}\right)\bigg],
 \end{split}
\end{equation}
where for $u\leq k,$
\begin{equation*}
p_u =
\frac{\frac{u-1}{k(k-1)}\sum_{d=1}^D \Omega_d d (d-1)\frac{ \left[
\begin{array}{c}
k-u\\
d-2\\
\end{array}
\right]} {\left[
\begin{array}{c}
k-2\\
d-2\\
\end{array}
\right]}}{1 - u \sum_{d=1}^D \Omega_d d \frac{\left[
\begin{array}{c}
k-u\\
d-1\\
\end{array}
\right]}{\left[
\begin{array}{c}
k\\
d\\
\end{array}
\right]} - \sum_{d=1}^D \Omega_d \frac{\left[
\begin{array}{c}
k-u\\
d\\
\end{array}\right]}{\left[
\begin{array}{c}
k\\
d\\
\end{array}\right]}},
\end{equation*}

and \[ \left[
\begin{array}{c}
a\\
b\\
\end{array}\right] := \binom{a}{b} b!,\] and $p_{k+1}:=\Omega_1.$ Further, $P_{k+1}(x,y):=x^n.$\\
\end{theorem}

This recursion gives a way to compute the probability of a decoding
error at each step of the BP decoding as 
\[P_{err}(u) = \sum_{c\geq 0} p_{c,0,u} = 1 - \sum_{c \geq 0, r \geq 1} p_{c,r,u} = 1 - P_u(1,1),\]
and the overall error probability of the decoder as
\[P_{err} = \sum_{u=1}^{k} P_{err}(u).\]

If we approximate the LT process by allowing output symbols to choose
their neighbors with replacement during encoding, $p_u$ becomes: 
\begin{equation*}
 p_u = \frac{1}{k}f\left(\frac{u}{k}\right) - \frac{1}{k^2}g\left(\frac{u}{k}\right) =  \frac{1}{k}f\left(\frac{u}{k}\right) +O(1/k^2),
\end{equation*}
where
\[f(x) := \frac{x \Omega''(1-x)}{1 - x \Omega{'}(1-x) - \Omega(1-x)} \mbox{ and } g(x):= \frac{f(x)}{x}.\]

With this assumption, Karp et al. use the recursion to derive
difference equations for the expected size of the ripple and the
cloud, and further approximate these difference equations by
differential equations that they solve to get closed-form expressions
for the expected ripple and cloud size. Formally, let $R(u)$ denote
the expected number of output symbols in the ripple, and $C(u)$ denote
the expected number of output symbols in the cloud, when $u$ input
symbols are undecoded, where $u$ is assumed to be a constant fraction
of the total number of input symbols $k$. Then the following theorem
shows that $R(u)$ is linear in $k$ for an appropriate choice of the LT
code parameters.\\

\begin{theorem}\cite{finite}\label{exp}
Consider an LT-code with parameters $(k,\Omega(x))$ and assume 
$n=(1+\epsilon)k$ symbols have been collected for decoding. During BP decoding, let $C(u)$ and $R(u)$ be respectively
the expected size of the cloud and ripple as a function of the number
$u$ of undecoded input symbols. Then, under the assumptions that $u$ is a constant fraction of $k$ and $\Omega_1 > 0,$ we have
 \begin{eqnarray*}
  C(u) &=& n\left(1-\frac{u}{k}\Omega'(1-u/k) - \Omega(1-u/k)\right)+O(1)\\
  R(u) &=& (1+\epsilon)u\left(\Omega'(1-u/k) + \frac{1}{1+\epsilon}\ln\frac{u}{k}\right)+O(1).
 \end{eqnarray*}
\end{theorem}

In what follows, we let $\hat{C}(x)$ be a continuous approximation of
$C(u/k):=C(u)/n,$ a normalized version of $C(u).$ $\hat{C}(x)$ can be
shown to be the solution of the differential equation 
\[\hat{C}'(x) = f(x) \hat{C}(x)\]
with initial condition \[\hat{C}(1)=C(1)=
(1-\Omega_1)\left(1-(1-\Omega_1)^{n-1}\right),\] and is given by 
\begin{equation*}
 \hat{C}(x) = c_0 \left(1-x\Omega'(1-x) - \Omega(1-x)\right),
\end{equation*}
with $c_0 = 1-(1-\Omega_1)^{n-1}.$

Similarly, we define $\hat{R}(x)$ as a continuous approximation of
$R(u/k):= R(u)/n.$ $\hat{R}(x)$ is the solution of 
\begin{equation*}
  \hat{R}'(x) = \frac{\hat{R}(x)}{x}  - c_0 x\Omega''(1-x)+\frac{1}{1+\epsilon}
\end{equation*}
with initial condition
 \[\hat{R}(1)=R(1)=\Omega_1 - \frac{1-(1-\Omega_1)^{n}}{n},\] and is given by
\begin{equation}\label{R_hat}
 \hat{R}(x) = x\left(c_0\Omega'(1-x) + \frac{1}{1+\epsilon}\ln{x}+r_0\right),
\end{equation}
with \[r_0 = \Omega_1(1-\Omega_1)^{n-1} - \frac{1-(1-\Omega_1)^n}{n}.\]

Then we can write
\begin{equation}\label{R}
 \begin{split}
  C(u) &=n\hat{C}(u/k) + O(1)\\
  R(u) &= n\hat{R}(u/k) + O(1).
 \end{split}
\end{equation}

\section{An Expression for the Variance of the Ripple Size}\label{var}
Let $\sigma_R^2(u)$ be the variance of the ripple size as a function
of the number of undecoded symbols $u$. In what follows we will always
assume that $u$ is a constant fraction of $k$. $\sigma_R^2(u)$ is
given by 
\begin{equation}\label{varR}
 \begin{split}
  \sigma_R^{2}(u) &= \sum_{c\geq 0, r \geq 1}(r-1)^2p_{c,r,u} - R(u)^2\\
&= N(u) - R(u)^2 + R(u),
 \end{split}
\end{equation}
where we define
\begin{equation}\label{defN}
 N(u) :=  \frac{\partial^2 P_u}{\partial y^2}(1,1)= \sum_{c \geq 0, r \geq 1}(r-1)^2 p_{c,r,u} - R(u).
\end{equation}
It is thus enough to find an expression for $N(u)$ to get an
expression for $\sigma_R^2(u).$ We start by differentiating both sides
of the recursion (\ref{recursion}) twice with respect to $y$ and
evaluating at $(1,1).$ This gives us a recursion for $N(u):$ 
\begin{equation}\label{N_rec}
 \begin{split}
N(u-1) &= \left(1 - \frac{1}{u}\right)^2 N(u)-2 p_u C(u) - 2 \left(1 -
\frac{1}{u}\right) R(u) \\ 
&+ p_u^2 \frac{\partial^2 P_u}{\partial x^2}(1,1) + 2 p_u\left(1 -
\frac{1}{u}\right)\frac{\partial^2 P_u}{\partial x \partial y}(1,1)\\ 
& - 2 \left[ - P_u(1,1)
+ P_u\left(1-p_u, \frac{1}{u}\right)\right].
\end{split}
\end{equation}
Before we can proceed with solving this difference equation, we need
to find expressions for the second-order derivatives $\frac{\partial^2
  P_u}{\partial x^2}(1,1)$ and $\frac{\partial^2 P_u}{\partial x
  \partial y}(1,1).$ We do so by following exactly the same method
that we are currently outlining for an expression for $N(u).$ Define 
\begin{eqnarray*}
 M(u) &:=& \frac{\partial^2 P_u}{\partial x^2}(1,1)\\
 L(u) &:=& \frac{\partial^2 P_u}{\partial x \partial y}(1,1).
\end{eqnarray*}
Let $\hat{M}(x)$ be a continuous approximation of the normalized
function $M(u/k):= M(u)/n^2.$ It can be shown that $\hat{M}(x)$ is the
solution of the differential equation 
\[\hat{M}'(x)=2 f(x)\hat{M}(x)\]
with initial condition
\[\hat{M}(x=1) = \left(1-\frac{1}{n}\right)(1 - \Omega_1)^2 \left(1 - (1-\Omega_1)^{n-2}\right),\]
and is given by the expression
\begin{equation*}
\hat{M}(x) = m_0 \left(1 - x \Omega'(1-x) - \Omega(1-x)\right)^2
\end{equation*}
with
\[m_0 = \left(1-\frac{1}{n}\right)\left(1 - (1-\Omega_1)^{n-2}\right).\]
Similarly, let $\hat{L}(x)$ be a continuous approximation of $L(u/k)
:= L(u)/n^2.$ It is the solution of 
\[ \hat{L}'(x) = \left(\frac{1}{x}+f(x)\right)\hat{L}(x) - f(x)\hat{M}(x) + \frac{1}{1+\epsilon}\hat{C}(x)\]
with initial condition
\[\hat{L}(x=1) = \left(1-\frac{1}{n}\right)\Omega_1(1-\Omega_1),\]
and an expression for it is
 \begin{equation*}
\begin{split}
\hat{L}(x) &= x\left(1-x\Omega'(1-x)-\Omega(1-x)\right)\\
&\qquad \cdot \left(m_0\Omega'(1-x)+\frac{c_0}{1+\epsilon}\ln{x}+l_0\right)
\end{split}
\end{equation*}
with
\[l_0 = \left(1-\frac{1}{n}\right)\Omega_1(1-\Omega_1)^{n-2} .\]
Then the following theorem gives closed-form expressions for $M(u)$ and $L(u).$\\

\begin{theorem}\label{ml}
\begin{eqnarray*}
  M(u) &=&n^2 \hat{M}(u/k)  + O(k)\\
L(u) &=& n^2 \hat{L}(u/k) +O(k).
\end{eqnarray*}
\end{theorem}

As for the ``dirt'' term
\begin{equation}\label{drift}
 - 2 \left[ - P_u(1,1)+ P_u\left(1-p_u, \frac{1}{u}\right)\right],
\end{equation}
it does not involve derivatives and we cannot use the same method to
find an expression for it independant the state generating
function. However, we can bound it under an assumption on the ripple
size. More specifically, it is not difficult to prove that 
for $r \geq 3$, the dirt term is of constant order. In what follows,
we assume that the size of the ripple does not go below the constant
$3$. 

Replacing $M(u)$ and $L(u)$ by their expressions and bounding the dirt
term in the recursion (\ref{N_rec}), we obtain the following
difference equation for $N(u):$ 

\begin{equation}\label{N_difference}
\begin{split}
N(u) &- N(u-1) = \left(\frac{2}{u}-\frac{1}{u^2}\right)N(u) - p_u^2 M(u) \\
&- 2 p_u\left(1 - \frac{1}{u}\right)L(u) +2 p_u C(u)\\
& + 2 \left(1 - \frac{1}{u}\right) R(u) + O(1).
\end{split}
\end{equation}

Note that $N(u)$ as defined in equation (\ref{defN}) can be as large
as a constant fraction of $k^2$. We thus need to normalize $N(u)$ if
we want to say something meaningful about the difference
$N(u)-N(u-1).$ We define $x:=u/k$ to be the \textit{fraction} of
undecoded symbols, and let $N(x) := N(u)/n^2$ be a normalized version
of $N(u).$ We similarly normalize the other functions of $u$ and
represent them as functions of $x$: 
\begin{eqnarray*}
M(x):=M(u)/n^2&,& L(x):= L(u)/n^2,\\
 C(x) := C(u)/n&,& R(x) := R(u)/n.
\end{eqnarray*}
Normalizing equation (\ref{N_difference}) and replacing the functions
$M(x),L(x),C(x)\mbox{ and }R(x)$ by their continuous approximations,
we obtain 
\begin{equation*}
\begin{split}
N(x) &- N(x-1/k) = \frac{2}{kx}N(x) -\frac{2}{k}f(x)\hat{L}(x)\\
&\qquad + \frac{2}{(1+\epsilon)k}\hat{R}(x)+ O(1/k^2).\\
\end{split}
\end{equation*}
Neglecting lower-order terms, we approximate $N(x)$ by the function $\tilde{N}(x)$ which satisfies
\begin{equation*}
\tilde{N}(x) - \tilde{N}(x-1/k) =\frac{2}{kx}\tilde{N}(x) -\frac{2}{k}f(x)\hat{L}(x) + \frac{2}{(1+\epsilon)k}\hat{R}(x),\\
\end{equation*}
with initial condition $\tilde{N}(1)=N(1).$\\
\begin{claim}\label{claim1}
For any $x$ on which $N(x)$ is defined, $N(x)$ and $\tilde{N}(x)$ differ by a term of the order of $1/k.$\\
\end{claim}

We skip the proof of this and subsequent claims for reasons of space, and refer the reader to the final version of this paper.

We further approximate the discrete function $\tilde{N}(x)$ by the continuous function $\hat{N}(x),$ and
\[\frac{\tilde{N}(x) - \tilde{N}(x-1/k)}{1/k}\]
by the first-order derivative of $\hat{N}(x)$. $\hat{N}(x)$ satisfies the differential equation
\begin{equation}\label{diffNhat}
 \hat{N}'(x) = \frac{2}{x} \hat{N}(x) - 2f(x)\hat{L}(x)+ \frac{2}{1+\epsilon}\hat{R}(x)
\end{equation}
with initial condition $\hat{N}(1) = \tilde{N}(1).$\\
\begin{claim}\label{claim2}
For any $x$ on which $\tilde{N}(x)$ is defined, $\tilde{N}(x)$ and $\hat{N}(x)$ differ by a term of the order of $1/k.$\\
\end{claim}

The general solution of the differential equation (\ref{diffNhat}) is given by
\begin{equation}\label{Nhat}
\begin{split}
 \hat{N}(x) &= x^2\Big(m_0\Omega'(1-x)^2 +2l_0\Omega'(1-x)+ \\
&\frac{2c_0}{1+\epsilon}\Omega'(1-x)\ln{x} + \frac{2r_0}{1+\epsilon}\ln{x}+ \frac{1}{(1+\epsilon)^2}(\ln{x})^2+ n_0\Big),
 \end{split}
\end{equation}
where the value of the constant $n_0$ can be found to be, by the initial conditions,
\[n_0 = -\left(1-\frac{1}{n}\right)(1-\Omega_1)^{n-2}\Omega_1^2 -\frac{2}{n}\Omega_1 + \frac{2}{n^2}\left(1-(1-\Omega_1)^n\right).\]
By claims \ref{claim1} and \ref{claim2} we thus have
\[N(x) = \hat{N}(x) +O(1/k),\]
where $\hat{N}(x)$ is given by equation (\ref{Nhat}). This gives us an expression for $N(u),$ up to a term of the order of $k$:
\begin{equation*}
\begin{split}
N(u)&=(1+\epsilon)^2u^2\bigg(\Omega'(1-u/k)^2+\frac{2}{1+\epsilon}\Omega'(1-u/k)\ln{\frac{u}{k}}\\
&\qquad+\frac{1}{(1+\epsilon)^2}\left(\ln{\frac{u}{k}}\right)^2\bigg)+O(k).
\end{split}
\end{equation*}
Comparing this expression to that for $R(u)^2$ given by equations (\ref{R_hat}) and (\ref{R}), it is easy to see that these two expressions agree up to terms of the order of $k$, so that the variance of the ripple size
\[\sigma_R^2(u) = N(u) - R(u)^2 + R(u)\]
is of the order of $k$.\\

\begin{theorem}
Consider an LT-code with parameters $(k,\Omega(x))$ and let $\sigma_R(u)$ be the standard deviation of the ripple size throughout BP decoding. Then
\[\sigma_R(u) = O(\sqrt k).\]
\end{theorem}
\section{Toward a Finite-Length Analysis of the LT Decoder}\label{finitelength}

Our ultimate goal is to be able to bound the error probability of the decoder as a function of $k$, without the assumption that $k$ goes to infinity. We thus need to find an expression for the variance of the ripple size, instead of simply determining its order. For this purpose, we must find an expression for $N(u)$ up to terms of constant order, and an expression for $R(u)$ up to terms of the order of $1/k.$ We illustrate the analysis for $N(u).$ From the recursion given by equation (\ref{N_rec}), we proceed by first, assuming that the ripple size does not go below $4$ so that the ``dirt'' term is of the order of $1/k$; and second, replacing $C(u),$ $R(u),$ $M(u),$ and $L(u)$ by finer approximations as follows:
\begin{equation}\label{precise_vals}
\begin{split}
 C(u)&= n\hat{C}(u/k) -n D_C(u/k)\\
 R(u)&= n\hat{R}(u/k) -n D_R(u/k) \\
M(u) &=n^2 \hat{M}(u/k) - n^2 D_M(u/k) \\
 L(u) &= n^2 \hat{L}(u/k) - n^2 D_L(u/k),
\end{split}
\end{equation}
where $D_C(x)$ is a discrepancy term introduced by approximating $C(u)$ by $\hat{C}(u),$ and $D_R(x), D_M(x) \mbox{ and } D_L(x)$ are defined similarly. These discrepancy terms are all of the order of $1/k$ and are given by the following expressions.

\begin{eqnarray*}
D_C(x) &=& \frac{1}{k^2}\sum_{i=0}^{k(1-x)-1}C_i\prod_{j=i+1}^{k(1-x)-1}\left(1-\frac{c_j}{k}\right)+O(1/k^2)\\
D_R(x) &=& \frac{1}{k^2}\sum_{i=0}^{k(1-x)-1}R_i\prod_{j=i+1}^{k(1-x)-1}\left(1-\frac{r_j}{k}\right)+O(1/k^2)\\
D_M(x) &=& \frac{1}{k^2}\sum_{i=0}^{k(1-x)-1} M_i\prod_{j=i+1}^{k(1-x)-1}\left(1-\frac{m_j}{k}\right)+O(1/k^2)\\
D_L(x) &=& \frac{1}{k^2}\sum_{i=0}^{k(1-x)-1} L_i\prod_{j=i+1}^{k(1-x)-1}\left(1-\frac{l_j}{k}\right) + O(1/k^2)\\
\end{eqnarray*}
 where $C_i, R_i, M_i, L_i$ and $c_j, r_j, m_j, l_j$ are constants for most of the decoding process and are given by
\begin{eqnarray*}
\textstyle C_i &=& \hat{C}''(1-i/k)-g(1-i/k)\hat{C}(1-i/k)\\
\textstyle c_j &=& f(1-j/k)\\
& &\\
R_i &=&\hat{R}''(1-i/k)+g(1-i/k)\hat{C}(1-i/k)\\
& & \,+k f(1-i/k) D_C(1-i/k)\\
r_j &=&\frac{1}{1-j/k}\\
& &\\
 M_i &=&  \hat{M}''(1-i/k)-\left(2g(1-i/k)+f(1-i/k)^2\right)\hat{M}(1-i/k)\\
 m_j &=&  2 f(1-j/k)\\
 & &\\
 L_i &=& \hat{L}''(1-i/k)-2g(1-i/k)\hat{L}(1-i/k)\\
& & \quad+\left(g(1-i/k)+f(1-i/k)^2\right)\hat{M}(1-i/k)\\
& & \quad + k f(1-i/k)D_M(1-i/k)\\
& & \quad -\frac{1}{1+\epsilon}f(1-i/k)\hat{C}(1-i/k)-\frac{k}{1+\epsilon}D_C(1-i/k)\\
& &\\
l_j &=& \frac{1}{1-j/k}+f(1-j/k).
\end{eqnarray*}

These expressions are obtained by the same method that we are now following to obtain a more precise approximation of $N(u).$

The next step is to write a recursion for $N(x)$ which is exact up to terms of the order of $1/k^3.$ We then approximate $N(x)$ by $\tilde{N}(x)$ which satisfies the same recursion except that we neglect terms of the order of $1/k^3$:
\begin{equation*}
\begin{split}
\tilde{N}(x) &- \tilde{N}(x-1/k) = \left(\frac{2}{kx}-\frac{1}{k^2x^2}\right)\tilde{N}(x) -  \frac{1}{k^2}f(x)^2 \hat{M}(x)\\
&+\left(-\frac{2}{k}f(x)+\frac{4}{k^2}g(x)\right)\hat{L}(x) + \frac{2}{k}f(x)D_L(x)\\
&+ \frac{2}{(1+\epsilon)k^2}f(x)  \hat{C}(x) + \left(\frac{2}{(1+\epsilon)k} - \frac{2}{(1+\epsilon)k^2x}\right) \hat{R}(x)\\
& -\frac{2}{(1+\epsilon)k}D_R(x)-\frac{2}{(1+\epsilon)^2k^2}.\\
\end{split}
\end{equation*}

\begin{claim}\label{claim3}
For any $x$ on which $N(x)$ is defined, $N(x)$ and $\tilde{N}(x)$ differ by a term of the order of $1/k^2.$\\
\end{claim}

We further approximate $\tilde{N}(x)$ by $\hat{N}(x)$ which satisfies the differential equation (\ref{diffNhat}) and is given by expression (\ref{Nhat}). A more careful analysis of the discrepancy beween $\hat{N}(x)$ and $\tilde{N}(x)$ leads to the following claim:\\

\begin{claim}\label{claim4}
For any $x$ on which $\tilde{N}(x)$ is defined, $\tilde{N}(x)$ and $\hat{N}(x)$ differ by a term of the order of $1/k.$\\ More precisely,
\[\hat{N}(x) - \tilde{N}(x) = D_N(x),\]
where
\begin{equation*}
 \begin{split}
  D_N(x)&= \frac{1}{k^2}\sum_{i=0}^{k(1-x)-1}\Bigg[\hat{N}''(1-i/k)-\frac{1}{(1-i/k)^2}\hat{N}(1-i/k)\\
&-f(1-i/k)^2 \hat{M}(1-i/k)+4g(1-i/k)\hat{L}(1-i/k)\\
&+2kf(1-i/k)D_L(1-i/k)+\frac{2f(1-i/k) }{(1+\epsilon)} \hat{C}(1-i/k) \\
&-\frac{2}{(1+\epsilon)(1-i/k)} \hat{R}(1-i/k)- \frac{2k}{1+\epsilon}D_R(1-i/k)\\
&-\frac{2}{(1+\epsilon)^2}\Bigg] \cdot \prod_{j=i+1}^{k(1-x)-1}\left(1-\frac{2}{k(1-j/k)}\right)+O(1/k^2).
 \end{split}
\end{equation*}
\end{claim}

By claims \ref{claim3} and \ref{claim4} we thus have
\[N(x) = \hat{N}(x) - D_N(x) + O(1/k^2),\]
where $\hat{N}(x)$ is given by equation (\ref{Nhat}). Using the resulting expression for $N(u)$, and the expression for $R(u)$ given by equation (\ref{precise_vals}), we finally get an expression for the variance of the ripple size up to terms of constant order.\\

\begin{theorem}
Consider an LT-code with parameters $(k,\Omega(x))$ and overhead $\epsilon$ and let $\sigma_R^2(u)$ be the variance of the ripple size throughout BP decoding. Then
\begin{equation*}
\begin{split}
\sigma_R^2(u) &=-(1+\epsilon)\frac{u^2}{k}\left(\Omega'(1-u/k)^2+ \frac{2}{1+\epsilon}\ln{\frac{u}{k}}+2\Omega_1\right)\\
&\quad\,+ (1+\epsilon)u\left(\Omega'(1-u/k) + \frac{1}{1+\epsilon}\ln{\frac{u}{k}}\right)\\
&\qquad \qquad \quad \cdot \left(1+2\left(\frac{u}{k}+nD_R(u/k)\right)\right)\\
&\quad\,-n^2D_N(u/k)+O(1).
\end{split}
\end{equation*}
\end{theorem}

%\section{An Example}

\begin{figure}
\includegraphics[totalheight=0.3\textheight,width=0.5\textwidth]{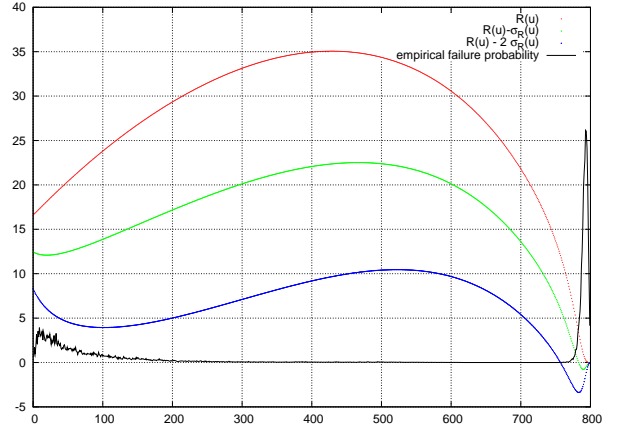}
\caption{Ripple size expectation and standard deviation versus the fraction of decoded input symbols. The black line is the empirical failure probability of the decoder based on 100 million simulations. It confirms that the ``problem zones'' of the decoder are the ones predicted by the second moment method.}\label{CapSol}
\end{figure}

Figure \ref{CapSol} shows a plot of the expected ripple size and the
functions $h_1(u)$ and $h_2(u)$ given by equation (\ref{hc}),
throughout the decoding process, for an LT-code with $k=800$ and
$\epsilon=0.1,$ and with the ``Capped Soliton'' degree distribution 
\[ \Omega(x)=\frac{1}{\frac{1}{50}+
  \sum_{i=2}^{50}\frac{1}{i(i-1)}}\left[\frac{1}{50}x +
  \sum_{i=2}^{50}\frac{1}{i(i-1)}x^i\right],\] 
inspired from Luby's Ideal Soliton distribution \cite{luby}. The plot
also shows the result of real simulations of this code, and confirms
that the problem zones of the decoder are those predicted by the
functions $h_i(u)$: the closer they are to the $x$-axis, the more
probable it is that the decoder fails.
As can be seen, there is a fair chance that the decoder fails when the
fraction of decoded input symbols is between 0 and 0.2, and there is a
  very good chance that the decoder fails when the fraction of decoded
  input symbols is close to 0.95.

\section{Conclusion}
We have given an analytic expression for the variance of the ripple size throughout the LT decoding process. This expression is asymptotically of the order of $k$, and we have expressed it as a function of $k$ as a first step toward finite-length analysis of the LT decoding. The next step is to work around the assumption that $u$ is a ``constant fraction'' of $k$. Then we would obtain a guarantee for successful decoding as a function of the LT-code parameters and overhead for practical values of $k$. This would then allow us to solve the corresponding design problem, namely to choose degree distributions that would make the function $h_c(u)$ stay positive for as large a value of $c$ as possible, for a fixed code length $k$.

%\bibitem{Shannon1948}
%C. E. Shannon, ``A mathematical theory of communication,''
%\emph{Bell Syst.\ Tech.\ J.}, vol.\ 27, pt.~I, pp.~379--423, 1948;
%     pt.~II, pp.~623--656, 1948.

\end{document}